\documentclass[12pt,a4paper]{article}
\usepackage{amsmath}
\usepackage{bm}
\usepackage{amsfonts}
\usepackage{graphicx}
\usepackage[normal]{caption2}
\usepackage{subfigure}
\usepackage{rotating}
\usepackage{citesort}
\usepackage{lscape}
\usepackage{section}
\begin{document}
\renewcommand\arraystretch{1.1}
\setlength{\abovecaptionskip}{0.1cm}
\setlength{\belowcaptionskip}{0.5cm}
%%%%%%%%%%%%%%%%%
\title {Geometry of vanishing flow: a new probe to determine the in-medium nucleon nucleon cross-section}
%%%%%%%%%%%%%%%%%%%%%%%%%%%%%%%%
\author {Rajiv Chugh$^{1}$ and Aman D. Sood$^{2}$\\
\it $^{1}$Department of Physics, Panjab University, Chandigarh
-160 014, India.\\
$^{2}$SUBATECH, \\Laboratoire de Physique Subatomique et des
Technologies Associ\'{e}es\\
Universit\'{e} de Nantes - IN2P3/CNRS - EMN \\
4 rue Alfred Kastler, F-44072 Nantes, France.\\} \maketitle
\begin{abstract}
We study the transverse flow throughout the mass range from
$^{20}Ne+^{20}Ne$ to $^{131}Xe+^{131}Xe$ as a function of the
impact parameter. We find that at smaller impact parameters the
flow is negative while going through the impact parameter,
transverse flow vanishes at a particular colliding geometry named
GVF. We find that the mass dependence of GVF is insensitive to the
equation of state and momentum dependent interactions whereas it
is quite sensitive to the cross section. So it can act as a useful
tool to pin down the nucleon nucleon cross section.
\end{abstract}
%PACS number: 25.70.-z, 25.70.Jj \\
 Electronic address:~amandsood@gmail.com
\newpage
\section{Introduction}
The heavy-ion reactions at intermediate energies have been used
extensively over the last three decades to produce the hot and
dense nuclear matter leading to the understanding of nuclear
matter equation of state (EOS) as well as in-medium
nucleon-nucleon (nn) cross-section. Collective transverse in-plane
flow \cite{sch,gust,doss} has been found to be one of the most
sensitive observable in this direction. A lot of experimental as
well as theoretical efforts have been made to study the transverse
in-plane flow \cite{ogli,sun98}. The variation in the flow as a
function of beam energy reflects the competition between the
attractive and repulsive interactions \cite{zhang,hong}. At a
particular incident energy, the strength of these two interactions
counter-balance each other and the net transverse in-plane flow
vanishes. This energy is often referred as the energy of vanishing
flow (EVF). Krofcheck ${et. al.}$ \cite{krof89}, for the first
time, reported the vanishing of collective flow in the reaction of
$^{139}La$+$^{139}La$ at around 50 MeV/nucleon. Later, a large
number of attempts were made to study the EVF over wide range of
masses and impact parameters. A comparison with experimental data
also threw light on the equation of state
\cite{luk05,mota,west1,pak,mag,mag1,sood,sull,buta}. The EVF has
been found to depend also on the combined mass of the system
\cite{west1}. A power law mass dependence ($\propto A^{\tau}$) has
also been reported in the literature. From earlier measurements,
$\tau$ was supposed to be close to -1/3 (resulting from the
interplay between the attractive mean field and repulsive nn
collisions) \cite{west1} whereas recent attempts suggested a
deviation from the above mentioned power law \cite{mag,mag1,sood}.
\par
The colliding geometry also plays an important role in determining
the flow as well as its disappearance
\cite{luk05,sull,mag,buta,pak,sun98}. As two nuclei collide, the
pressure and density increase in the interaction region. At non
zero colliding geometry, due to the anisotropy in the pressure,
there is a transverse flow of nuclear matter in the direction of
lowest pressure. Therefore, as colliding geometry increases from
perfectly central collisions (i.e. b = 0), the transverse in-plane
flow increases, passes through a maximum and with further increase
in the impact parameter, the transverse flow decreases and passes
through a zero value and achieves even negative values \cite{pak}.
For grazing collisions, it must vanish again. Thus, barring the
perfectly central and grazing collisions, there should exist a
particular colliding geometry at a given incident energy at which
the transverse flow must vanish.
\par
In this letter, we propose a new observable, the geometry of
vanishing flow (GVF), which can be used to extract the information
about the EOS as well as about the in-medium nn cross-section.
Here, we shall show that the GVF depends on the combined mass of
the system and follows a power law behavior $\propto A^{\tau}$.
\par
For the present study, we use quantum molecular dynamics (QMD)
model. In QMD model \cite{aichqmd,qmd2,qmd3,qmd4}, each nucleon is
represented by a Gaussian distribution whose centroid propagates
with classical equations of motion:
\begin{equation}
\frac {d {\bf r}_i}{dt} = \frac{d H} {d{\bf p}_i};
\end{equation}
\begin{equation}
\frac{d{\bf p}_i}{dt} = - \frac{d H}{d{\bf r}_i},
\end{equation}
where the Hamiltonian is given by :
\begin{equation}
H = \sum_i \frac{{\bf p}_i^2}{2m_i} + V^{tot},
\end{equation}
with total interaction potential
\begin{equation}
V^{tot} = V^{Loc} + V^{Yuk} + V^{Coul}.
\end{equation}
Here $V^{Loc}$, $V^{Yuk}$ and $V^{Coul}$  represent, respectively,
the Skyrme, Yukawa and Coulomb parts of the interaction. Yukawa
force is known to be very important for low energy phenomena like
fusion, cluster decay etc. \cite{dutt}. Using the above mentioned
interactions, the static part of the mean field acting on each
nucleon can be written as:
\begin{equation}
V^{Loc} = \frac{\alpha}{2}({\frac{\rho}{\rho_{0}}}) +
\frac{\beta}{\gamma+1}({\frac{\rho}{\rho_{0}})^{\gamma}}.
\end{equation}
The parameters $\alpha$ and $\beta$ ensure right binding energy of
the colliding nuclei and $\gamma$ gives freedom to choose
different equations of state. The momentum dependent interactions
can be incorporated from the momentum dependence of the real part
of the optical potential.
%\begin{equation}
%\end{equation}
The final form of the momentum dependent potential ($V^{MDI}$) is
given as
\begin{equation}
V^{MDI} = \delta.ln^{2}[\varepsilon(\rho/\rho_{0})^{2/3}+
1]\rho/\rho_{0}.
\end{equation}
with $\delta$, $\varepsilon$ having values equal to 1.57 MeV and
21.54.
\begin{figure}
\begin{center}
\includegraphics[width=8cm]{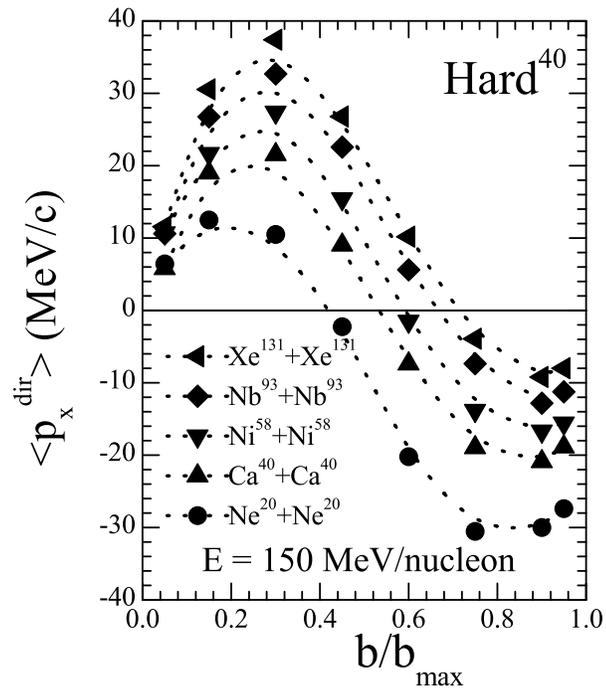}
%\vskip - 1.0cm
\caption{$<p_{x}^{dir}>$ (MeV/c) as a function of reduced impact
parameter $b/b_{max}$. We display the results of different systems
using hard equation of state and constant cross-section of 40 mb.}
\end{center}
\end{figure}
\par
For the present study, we simulated the reactions of
$^{20}Ne+^{20}Ne, ^{40}Ca+^{40}Ca, ^{58}Ni+^{58}Ni,
^{93}Nb+^{93}Nb$ and $^{131}Xe+^{131}Xe$ at full range of
colliding geometries ranging from the central to peripheral
collisions in small steps of 0.15 at a fixed incident energy of
150 MeV/nucleon. We used hard (dubbed as Hard), hard with momentum
dependent interactions (MDI)(labeled as HMD), and soft equation of
state with MDI (SMD) along with constant isotropic constant
cross-section of 40 mb as well as Cugnon parametrization of energy
dependent cross-section \cite{cug}. The superscripts to the labels
represent the different nn cross-sections. To calculate the
transverse in-plane flow, we use \textit{"directed transverse
momentum $<p_{x}^{dir}>$"} defined as \cite{sood,qmd2}
\begin{equation}
\langle{p_{x}^{dir}}\rangle = \frac{1}
{A}\sum_{i=1}^{A}{sign{{\{y(i)\}}p_{x}(i)}},
\end{equation}
where $y(i)$ and $p_{x}(i)$ are the rapidity and momentum of
$i^{th}$ particle, respectively.
\par
In fig.1, we display the $<p_{x}^{dir}>$ as a function of reduced
impact parameter $b/b_{max}$ (where $b_{max}$ = radius of
projectile + radius of target) for different colliding masses
between $^{20}Ne+^{20}Ne$ and $^{131}Xe+^{131}Xe$ at incident
energy of 150 MeV/nucleon. All reactions were followed till
$<p_{x}^{dir}>$ saturates. The saturation time is longer for
heavier masses compared to lighter ones. As expected, in all
cases, $<p_{x}^{dir}>$ increases with increase in the $b/b_{max}$
from perfectly central collisions and reaches a maximal value.
After maximal value, it decreases with further increase in the
colliding geometry, passes through a zero value at some
intermediate impact parameter. This colliding geometry at which
$<p_{x}^{dir}>$ passes through a zero value has been dubbed as
geometry of vanishing flow (GVF). With further increase in the
$b/b_{max}$, $<p_{x}^{dir}>$ becomes negative and attains the
maximal negative value after which it again vanishes at grazing
collisions. The trend is uniform through out the present mass
range. The value of the GVF varies with the mass of the combined
system. For lighter systems, the value of GVF is smaller compared
to the heavier ones.
\par
\begin{figure}
\centering
\begin{center}
\includegraphics[width=8cm]{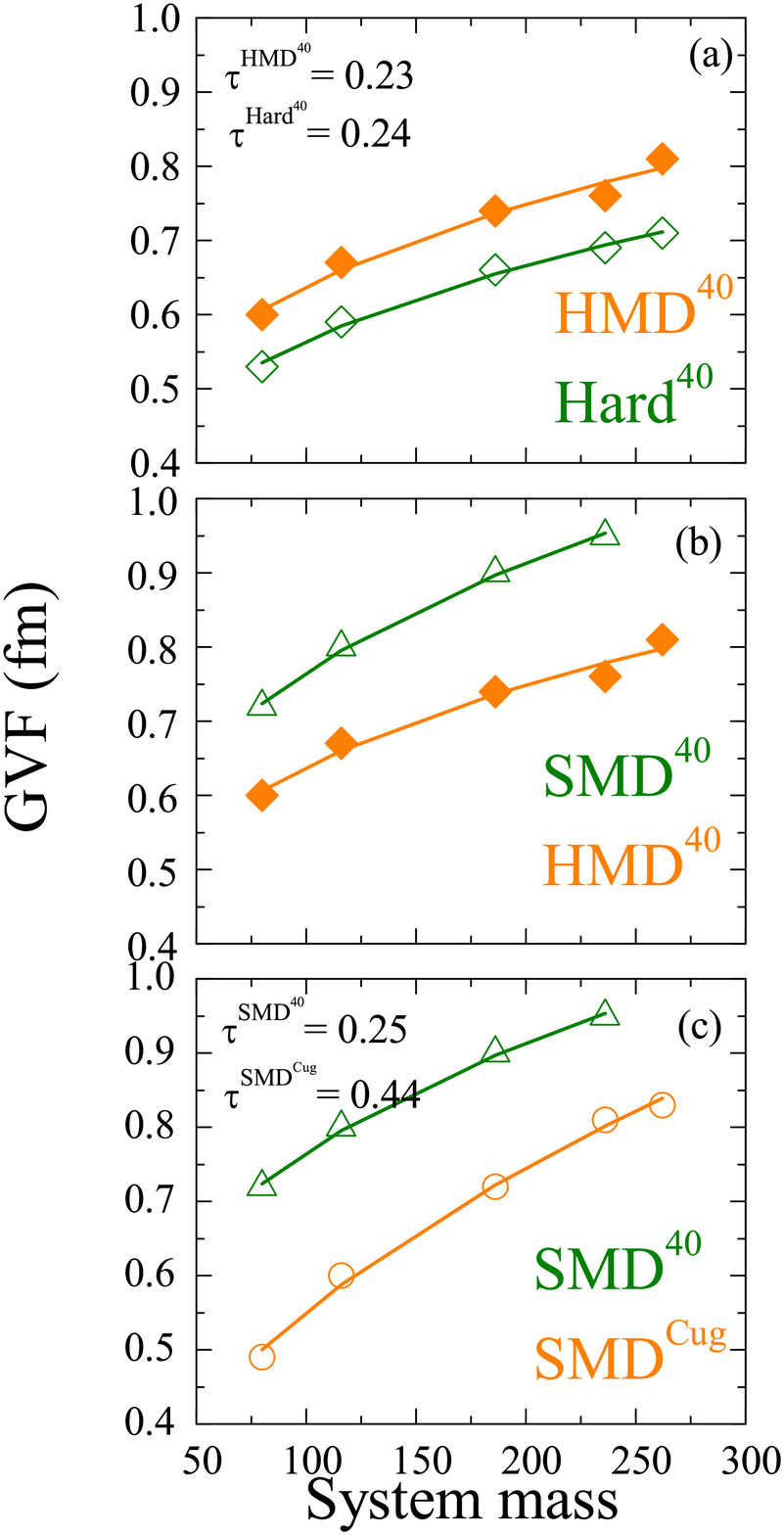}
%\vskip - 1.0cm
\caption{The geometry of vanishing of flow (GVF) as a function of
system mass for different equations of state.}
\end{center}
\end{figure}
In fig.2, we display GVF as a function of combined system mass. In
fig. 2a, we display the results of Hard$^{40}$ (open diamonds) and
HMD$^{40}$ (solid diamonds) EOS. The results of fig. 2b are for
HMD$^{40}$ and SMD$^{40}$ (open triangles) EOS whereas in fig. 2c,
we display the results for SMD$^{40}$ and SMD$^{Cug}$ (open
circles) EOS. The lines are power law fit ($\propto A^{\tau}$).
\begin{figure}
%\centering
\begin{center}
\includegraphics[width=14cm]{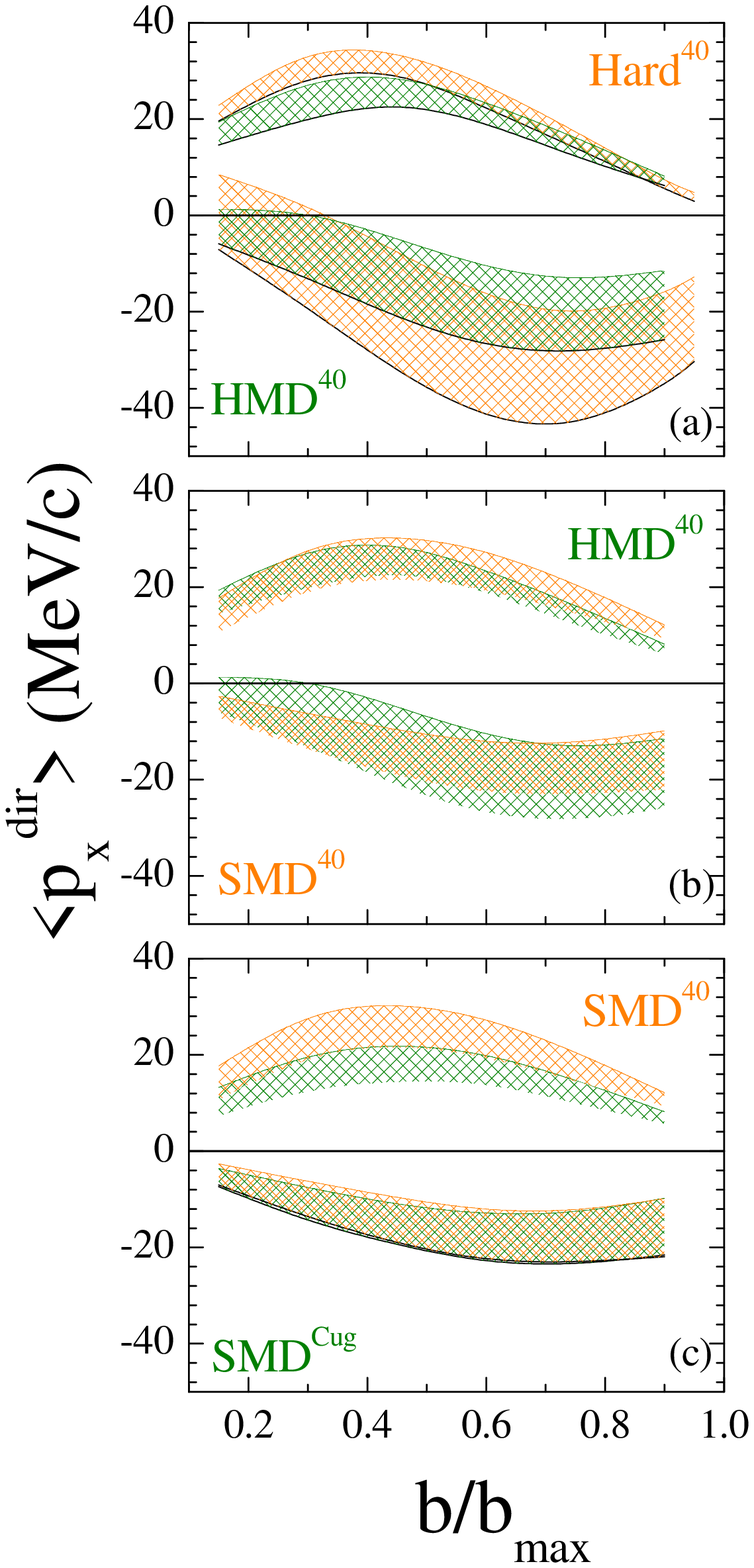}
%\vskip - 1.0cm
\caption{The decomposition of $<p_{x}^{dir}>$ into mean field and
binary collision parts as a function of the reduced impact
parameters for different equations of state.}
\end{center}
\end{figure}
In all the cases, GVF follows a power law behavior
$\propto{A^{\tau}}$, where A is the combined mass of the system.
The values of $\tau^{Hard^{40}}$, $\tau^{HMD^{40}}$,
$\tau^{SMD^{40}}$ and $\tau^{SMD^{Cug}}$ are, respectively, 0.24
$\pm$ 0.01, 0.23 $\pm$ 0.02, 0.25 $\pm$ 0.01 and 0.44 $\pm$ 0.02.
Note that the values of $\tau^{SMD^{Cug}}$ is quite different from
the values of $\tau^{Hard^{40}}$, $\tau^{HMD^{40}}$ and
$\tau^{SMD^{40}}$ which are almost same, indicating that the mass
dependence of GVF is insensitive towards the different equations
of state as well as towards the momentum dependent interactions.
It is, however, very sensitive towards the in-medium nn
cross-section. Note that the average strength of $\sigma^{Cug}$ in
Fermi energy is about 32 mb and with a slight change in the
cross-section, the value of $\tau$ is almost doubled. Therefore,
the mass dependence of GVF can be used to pin down the in-medium
nn cross-section. Moreover, it can also be used to explore the
isospin effects related to the cross-section since np cross
section is about a factor of three higher than the nn or pp
cross-section in the present energy range.
\par
As reported in Refs. \cite{sun98,sood}, the variation in the
strength of nucleon-nucleon cross-section by keeping the form of
all cross-sections same, yields a linear variation in the
collective flow. In Refs. \cite{sun98,sood}, different strengths
of cross-sections (eg. $\sigma$ = 20, 35, 40 and 55 mb) was used
and collective flow was found to vary linearly in agreement with
other calculations \cite{xu}.
\par
To understand the above behavior, we divide the total
$<p_{x}^{dir}>$ into the contribution from the mean field and
binary nn collision flow. The decomposition of $<p_{x}^{dir}>$ is
explained at energy of vanishing flow in Ref. \cite{sood}. In fig.
3, we display the $<p_{x}^{dir}>$ due to mean field (labeled as
$<p_{x}^{dir}>_{mf}$ and collisions ($<p_{x}^{dir}>_{coll}$) as a
function of the colliding geometry. In the present study,
$<p_{x}^{dir}>_{coll}$ is always positive where as
$<p_{x}^{dir}>_{mf}$ is always negative except at very small
colliding geometry (which is not relevant in context of the
present study). The shaded areas cover the full range of mass in
the present study. The values of both $<p_{x}^{dir}>_{mf}$ and
$<p_{x}^{dir}>_{coll}$ are larger for heavier mass than the
lighter ones i.e. upper (lower) boundary of each shaded area
represent heavier (lighter) mass. The top panel is for $Hard^{40}$
and $HMD^{40}$, bottom panel is for $SMD^{40}$ and $SMD^{Cug}$ and
middle panel is for $SMD^{40}$ and $HMD^{40}$. From figs. 2a and
3a, we see that the range of GVF (from lighter to heavy masses) is
around 0.5 to 0.8 and in this range, the inclusion of MDI has the
same effect on GVF throughout the mass range keeping the value of
$\tau$ unchanged. Similar effects can also be seen for the
equation of state from figs. 2b and 3b. From figs. 2c and 3c, we
see that flow due to the mean field is exactly the same for all
the systems at all colliding geometries, whereas, the effect of
binary collisions is different for different masses at and around
respective GVF, thus making the mass dependence of GVF quite
sensitive to the nucleon-nucleon cross-section.
\par
%Summary
In summary, we have studied the transverse flow throughout the
mass range from $^{20}Ne+^{20}Ne$ to $^{131}Xe+^{131}Xe$ as a
function of impact parameter. We find that at smaller impact
parameters, the flow is negative while going through the impact
parameter, transverse flow vanishes at a particular colliding
geometry dubbed as the GVF. We find that the mass dependence of
GVF is insensitive to the equation of state and momentum dependent
interactions whereas it is quite sensitive to the binary nn
cross-section presenting it a new probe to constraint the strength
of binary cross-section.

This work is supported by Indo-French Center for the Promotion of
Advanced Research (IFCPAR), New Delhi, project no. IFC/4104-1.
%%%%%%%%%%%%%%%%%%%%%%%%%%%%%%%%%%%%%%%%%%%%%%%%%%%%%%%%%%%%%%%%%%%%%%%


\begin{thebibliography}{999}
\bibitem{sch} W. Scheid, H. Muller and W. Greiner, {\it Phys. Lett.} {\bf 32,}
741 (1974).
\bibitem{gust} H. {\AA}. Gustafson \emph{et al.}, {\it Phys. Lett.} {\bf 52,}
1590 (1984).
\bibitem{doss} S. Gautam {\it et. al.} {\it J. Phys. G} {\bf 37,}
085102 (2010); Y. K. Vermani {\it et. al.} {\it Nucl. Phys. A}
{\it et. al.} {\bf 847,} 243 (2010).
\bibitem{ogli} C. A. Ogilvie \emph{et al.}, {\it Phys. Rev. C} {\bf 40,}
2592 (1989); B. Bl\"{a}ttel \emph{et al.}, {\it Phys. Rev. C} {\bf
43,} 2728 (1991); A. Andronic \emph{et al}., {\it Phys. Rev. C}
\textbf{67}, 034907 (2003).

\bibitem{sun98} S. Kumar \emph{et al.}, {\it Phys. Rev. C} \textbf{58},
3494 (1998); E. Lehmann, {\it Z. Physik A} \textbf{355}, 55
(1996); A. D. Sood and R. K. Puri, {\it Phys.Rev. C} {\bf 70,}
034611 (2004); S. Kumar \emph{et al.}, {\it Phys. Rev. C}
\textbf{81}, 014601 (2010); R. Chugh and R. K. Puri {\it Phys.
Rev. C} {\bf 82}, 014603 (2010).
%\bibitem{bla91}
%\bibitem{ram} V. Ramillien \emph{et al.}, Nucl. Phys. A {\bf 587,}
%802 (1995).
%\bibitem{wmzhang} W. M. Zhang \emph{et al.} Phys. Rev. C {\bf 42,}
%R491 (1990), M. D. Partlan \emph{et al.} Phys. Rev. Lett. {\bf
%75,} 2100 (1995), P. Crochet \emph{et al.} Nucl. Phys. {\bf A624,}
%755(1997).
%\bibitem{gale} C. Gale \emph{et al.}, Phys. Rev. C {\bf 41,} 1545 (1990).
%\bibitem{andro}
%\bibitem{pandan}
%\bibitem{ram95}
%\bibitem{bert87}
\bibitem{zhang} Y. Zhang and Z. Li, {\it Phys. Rev. C} \textbf{74}, 014602
(2006); W. M. Zhang \emph{et al.}, {\it Phys. Rev. C} {\bf 42,}
R491 (1990).
%\bibitem{wmz}
%\bibitem{beav}
%\bibitem{luk08}
\bibitem{hong} B. Hong \emph{et al.}, {\it Phys. Rev. C} \textbf{66}, 034901
(2000).
\bibitem{krof89} D. Krofcheck \emph{et al.}, {\it Phys. Rev. Lett.} \textbf{63,}
2028 (1989).
\bibitem{luk05} J. Lukasik \emph{et al.}, {\it Phys. Lett. B} {\bf
608,} 223 (2005).
\bibitem{mota} V. de la Mota \emph{et al.}, {\it Phys. Rev. C} {\bf 46,}
677(1992); D. Cussol \emph{et al.}, {\it Phys. Rev. C}  {\bf 65,}
044604 (2002); H. Zhou, Z. Li and Y. Zhuo, {\it Phys. Rev. C} {\bf
50,} R2664 (1994); {\it ibid} {\it Nucl. Phys.} {\bf A580,}
627(1994).

\bibitem{west1} G. D. Westfall \emph{et al.}, {\it Phys. Rev. Lett.} {\bf 71,}
1986 (1993).
\bibitem{pak} R. Pak {\it et al.}, {\it Phys. Rev. C} {\bf 53,}
1469 (1996).
%\bibitem{cuss}
%
%\bibitem{zhou}
%\bibitem{zhou1}
\bibitem{mag1} D. J. Magestro \emph{et al.}, {\it Phys. Rev. C}{\bf
61,} 021602(R) (2000).
\bibitem{mag} D. J. Majestro, W. Bauer and G. D. Westfall, {\it Phys. Rev. C} {\bf 62,}
041603 (2000).
\bibitem{sood} A. D. Sood and R. K. Puri, {\it Phys.Rev. C} {\bf 69,}
054612 (2004); {\it ibid} {\it Phys. Lett. B} {\bf 594,} 260
(2004); {\it ibid} {\it Euro. Phys. J. A} {\bf 30,} 571 (2006);
{\it ibid} {\it Phys. Rev. C} {\bf 73,} 067602 (2006).
\bibitem{sull} J. P. Sullivan {\it et al.}, {\it Phys. Lett. B} {\bf 249,}
8 (1990).
\bibitem{buta} A. Buta {\it et al.}, {\it Nucl. Phys.}  {\bf A584,}
397 (1995).
\bibitem{aichqmd} J. Aichelin, {\it Phys. Rep.} {\bf 202},
233 (1991), S. Kumar \emph{et al.}, {\it Phys. Rev. C} {\bf 78,}
064602 (2008).
\bibitem{qmd2} J. Singh \emph{et al.}, {\it Phys. Rev. C} {\bf
62}, 044617 (2000), R. K. Puri and J. Aichelin, {\it J. Comput.
Phys.} {\bf 162}, 245 (2000); Y. K. Vermani \emph{et al.}, {\it J.
Phys. G: Nucl. Part. Phys.} \textbf{37}, 015105 (2010); {\it ibid}
{\it Phys. Rev. C} \textbf{79}, 064613 (2009); Y. K. Vermani and
R. K. Puri, {\it Euro. Phys. Lett.} \textbf{85}, 62001 (2009); A.
D. Sood and R. K. Puri, {\it Phys. Rev. C} \textbf{79}, 064618
(2009).



\bibitem{qmd3} S. Kumar, R. K. Puri, and J. Aichelin, {\it Phys. Rev. C} \textbf{58}, 1618 (1998);
R. K. Puri, C. Hartnack, and J. Aichelin, {\it Phys. Rev. C} {\bf
54}, R28 (1996); G. Batko {\it et al.}, {\it J. Phys. G}{\bf
20},461 (1994); S. Huang \emph{et al.}, {\it Prog. Part. Phys.}
{\bf 30}, 105 (1993); C. Fuchs \emph{et al.}, {\it J. Phys. G}
{\bf 22,} 131 ( 1996); Y. K. Vermani \emph{et al.},{\it J. Phys.
G} {\bf 36,} 105103 (2009); S. Kumar \emph{et al.}, {\it Phys.
Rev. C} {\bf 81,} 014611 (2010);
\bibitem{qmd4} C. Hartnack \emph{et al.}, {\it Euro. Phys. J. A} {\bf 1,} 151 (1998);
A. Deckmyn, {\it Phys. Lett. B} {\bf 298}, 318 (1993); P. B.
Gossiaux {\it et al.}, {\it Nucl. Phys. A} {\bf 619,} 379 (1997);
R. K. Puri {\it et al.}, {\it Nucl. Phys. A} {\bf 575,} 733
(1994); E. Lehmann \emph{et al.}, {\it Phys. Rev. C} {\bf 51,}
2113 (1995); E. Lehmann {\it et al.}, {\it Prog. Part. Nucl.
Phys.} {\bf 30,} 219 (1993); S. Kumar and R. K. Puri, {\it Phys.
Rev. C} {\bf 58,} 320 (1998); {\it ibid} {\it Phys. Rev. C} {\bf
60,} 054607 (1999); {\it ibid} {\it Phys. Rev. C} {\bf 62,} 054602
(2000); {\it ibid} {\it Phys. Rev. C} {\bf 58,} 2858 (1998).
\bibitem{dutt} I. Dutt {\it et. al.}, {\it Phys. Rev. C} {\bf 81,}
064608 (2010); {\it ibid} {\bf 81,} 064609 (2010); {\bf 81 ,}
044615 (2010); {\bf 81,} 047601 (2010), R. K. Puri {\it et.
al.},{\it Euro. Phys. J. A} {\bf 23,} 429 (2005);{\it ibid} {\bf
8,} 103 (2000);{\it ibid} {\bf 3,} 277 (1998); {\it ibid} {\it
Phys. Rev. C} {\bf 45,} 1837 (1992); {\it ibid} {\bf 43,} 315
(1991);{\it ibid} {\it J. Phys. G} {\bf 18,} 903 (1992).
\bibitem{cug} J. Cugnon, T. Mizutani and J. Vandermeulen, {\it Nucl.
Phys.} {\bf A352,} 505 (1981).
\bibitem{xu}H. M. Xu {\it Phys. Rev. Lett.} {\bf 67,} 2769 (1991).
\end{thebibliography}
\end{document}